\documentclass[preprint,showpacs,preprintnumbers,amsmath,amssymb]{revtex4}

\def\undersim#1{\setbox9\hbox{${#1}$}{#1}\kern-\wd9\lower
    2.5pt \hbox{\lower\dp9\hbox to \wd9{\hss $_\sim$\hss}}}
\usepackage{amssymb}
\usepackage{amsmath}
\usepackage{graphicx}

\def\undersim#1{\setbox9\hbox{${#1}$}{#1}\kern-\wd9\lower
    2.5pt \hbox{\lower\dp9\hbox to \wd9{\hss $_\sim$\hss}}}

\def\mx{{\mathbf x}}
\def\my{{\mathbf y}}
\def\mz{{\mathbf z}}
\def\mv{{\mathbf v}}
\def\mr{{\mathbf r}}

\def\mr{{\mathbf r}}

\def\mk{{\mathbf k}}

\begin{document}

\title{Back-to-back correlations of boson-antiboson pairs for anisotropic
expanding sources}

\author{Yong Zhang$^1$}
\author{Jing Yang$^1$}
\author{Wei-Ning Zhang$^{1,\,2}$\footnote{wnzhang@dlut.edu.cn}}
\affiliation{$^1$School of Physics and Optoelectronic Technology, Dalian
University of Technology, Dalian, Liaoning 116024, China\\
$^2$Department of Physics, Harbin Institute of Technology, Harbin,
Heilongjiang 150006, China}

%\date{\today}

\begin{abstract}
In the hot and dense hadronic sources formed in high energy heavy ion
collisions, the particle interactions in medium might lead to a measurable
back-to-back correlation (BBC) of boson-antiboson pairs.  We calculate the
BBC functions of $\phi\,\phi$ and $K^+K^-$ for anisotropic expanding
sources.  The dependences of the BBC on the particle momentum and source
expanding velocity are investigated.  The results indicate that the BBC
functions increase with the magnitude of particle momentum and exhibit
an obvious dependence on the direction of the momentum for the anisotropic
sources.  As the source expanding velocity decreases, the BBC function
decreases when the particle momentum is approximately perpendicular to
the source velocity, and the BBC function increases when the particle
momentum is approximately parallel to the source velocity.


\end{abstract}

\maketitle

\section{Introduction}
In the mid to late 1990s, it was shown \cite{AsaCso96,AsaCsoGyu99} that the
mass modification of the particles in the hot and dense hadronic sources can
lead to a squeezed back-to-back correlation (BBC) of boson-antiboson pairs
in high energy heavy ion collisions.  This BBC is the result of a quantum
mechanical transformation relating in-medium quasiparticles to the two-mode
squeezed states of their free observable counterparts, through a Bogoliubov
transformation between the creation (annihilation) operators of the
quasiparticles and the free observable particles
\cite{AsaCso96,AsaCsoGyu99,Padula06}.  The investigations of the BBC may
provide another way for people to understand the thermal and dynamical
properties of the hadronic sources in high energy heavy ion collisions,
in addition to particle yields and spectra.

In Ref. \cite{Padula06}, S. Padula et al. put forward the formulism of the
BBC for the local-equilibrium expanding sources, and studied the BBC functions
of $\phi\phi$.  Recently, the BBC functions of $K^+K^-$ are also investigated
\cite{Padula10} based on the formulism \cite{Padula06}, and a method is
suggested \cite{Padula10a} to search for the squeezed BBC in the heavy
ion collisions at the Relativistic Heavy Ion Collider (RHIC) and the Large
Hadron Collider (LHC).
In Ref. \cite{YZHANG14}, we calculated the BBC functions of $\phi\phi$ and
$K^+K^-$ using a Monte Carlo method and investigated the relativistic effect
on the BBC functions.  However, all these previous works are performed for
spherical sources \cite{Padula06,Padula10,Padula10a,YZHANG14}.  Motivated by
the anisotropy of the hadronic sources formed in high energy heavy ion
collisions, we will study here the BBC functions for anisotropic hadronic
sources, using the relativistic Monte Carlo algorithm as in Ref. \cite{YZHANG14}.
We will investigate the dependences of the BBC functions on the anisotropic
source velocity and the angle between particle momentum and source velocity.
As compared to the early spherical source models, the momentum-direction
dependence of the BBC function for anisotropic sources provides additional
signals for experimental detection.

In the next section we will give a brief description of the BBC formulas
used in the calculations.  In Sec. III, we will present the results of the
BBC functions of $\phi\,\phi$ and $K^+K^-$ varying with particle momentum
and source velocity.  We will further examine the effect of source velocity
on the BBC functions in Sec. IV, and finally provide the summary and
conclusions in Sec. V.

\section{BBC functions for anisotropic expanding sources}
Denote $a_\mk\, (a^\dagger_\mk)$ as the annihilation (creation) operators
of the freeze-out boson with momentum $\mk$ and mass $m$, and $b_\mk\,
(b^\dagger_\mk)$ as the annihilation (creation) operators of the
corresponding quasiparticle with momentum $\mk$ and modified mass $m_{\!*}$
in the homogeneous medium, they are related by the Bogoliubov transformation
\cite{AsaCso96,AsaCsoGyu99}
\begin{equation}
a_{\mk} = c_{\mk}\,b_{\mk} + s^*_{-\mk}\,b^\dagger_{-\mk},
\vspace*{-2mm}
\end{equation}
where
\begin{equation}
c_{\mk} = \cosh f_{\mk}\,,~~~~~s_{\mk} = \sinh f_{\mk}\,,~~~~~
f_{\mk} = \frac{1}{2} \log (\omega_{\mk}/\Omega_{\mk}),
\end{equation}
\begin{equation}
\omega_\mk=\sqrt{\mk^2 + m^2}\,,~~~~~\Omega_\mk=\sqrt{\mk^2+m_{\!*}^2}\,.
\end{equation}
The BBC function is defined as \cite{AsaCso96,AsaCsoGyu99}
\begin{equation}
\label{BBCf}
C(\mk,-\mk) = 1 + \frac{|G_s(\mk,-\mk)|^2}{G_c(\mk,\mk) G_c(-\mk,-\mk)},
\end{equation}
where $G_c(\mk_1,\mk_2)$ and $G_s(\mk_1,\mk_2)$ are the chaotic and squeezed
amplitudes, respectively, as
\begin{equation}
G_c(\mk_1,\mk_2)=\sqrt{\omega_{\mk_1}\omega_{\mk_2}}\,\langle
a^\dagger_{\mk_1} a_{\mk_2} \rangle,
\end{equation}
\begin{equation}
G_s(\mk_1,\mk_2)=\sqrt{\omega_{\mk_1}\omega_{\mk_2} }\,\langle
a_{\mk_1} a_{\mk_2}\rangle,
\end{equation}
where $\langle \cdots \rangle$ means ensemble average.

For local-equilibrium expanding sources with the sudden freeze-out
assumption, $G_c(\mk_1,\mk_2)$ and $G_s(\mk_1,\mk_2)$ can be expressed
as \cite{AsaCsoGyu99,Sinyukov94,Padula06,Padula10,Padula10a,YZHANG14}
\begin{eqnarray}
\label{Gc}
&&G_c(\mk_1,\mk_2)=\frac{K^0_{1,2}\,{\widetilde F}(\omega_{\mk 1}\!\!
-\!\omega_{\mk 2})}{(2 \pi)^3}\!\!\int \!\!d^3 r\, e^{i(\mk_1-\mk_2)
\cdot\mr}\rho(\mr)\nonumber\\
&&\hspace*{18mm}\times\Big\{|c_{\mk1,\mk2}(\mr)|^2n_{\mk1,\mk2}(\mr)
+|s_{\mk1,\mk2}(\mr)|^2 [n_{\mk1,\mk2}(\mr)+1]\Big\},
\end{eqnarray}
\begin{eqnarray}
\label{Gs}
&&G_s(\mk_1,\mk_2)=\frac{K^0_{1,2}\,{\widetilde F}(\omega_{\mk 1}\!\!
+\!\omega_{\mk 2})}{(2 \pi)^3}\!\!\int \!\!d^3 r\,e^{i(\mk_1+\mk_2)
\cdot\mr}\rho(\mr)\nonumber\\
&&\hspace*{18mm}\times\Big\{s^*_{\mk1,\mk2}(\mr)\,c_{\mk2,\mk1}(\mr)
\,n_{\mk1,\mk2}(\mr)+c_{\mk1,\mk2}(\mr)\,s^*_{\mk2,\mk1}(\mr)
[n_{\mk1,\mk2}(\mr)+1] \Big\},
\end{eqnarray}
where
\begin{equation}
c_{\mk_1,\,\mk_2}(\mr)=\cosh[\,f_{\mk_1,\,\mk_2}(\mr)\,],~~~~~
s_{\mk_1,\,\mk_2}(\mr)=\sinh[\,f_{\mk_1,\,\mk_2}(\mr)\,],
\end{equation}
\begin{eqnarray}
f_{\mk_1,\,\mk_2}(\mr)=\frac{1}{2}\log\left[K^{\mu}_{1,2}u_{\mu}(\mr)
/K^{*\nu}_{1,2}u_{\nu}(\mr)\right],
\end{eqnarray}
\begin{eqnarray}
\label{nkk}
n_{\mk_1,\,\mk_2}(\mr)=\exp{\{-[K^{*\mu}_{1,2}u_\mu(\mr)-\mu_{1,2}(\mr)]
/\,T(\mr)\}},
\end{eqnarray}
where, $K^{\mu}_{1,2}=(k^{\mu}_1+k^{\mu}_2)/2$ and $K^{*\mu}_{1,2}=
(k^{*\mu}_1+k^{*\mu}_2)/2$ are the pair four-momenta of the particles
and the quasiparticles in medium, $u^{\mu}(\mr)=\gamma_\mv[1,\mv(\mr)]$,
$\mu_{1,2}(\mr)$, and $T(\mr)$ are the source four-velocity, the pair
chemical potential, and the source temperature at particle freeze-out,
respectively.  In Eqs. (\ref{Gc}) and (\ref{Gs}), ${\widetilde F}
(\omega_{\mk 1}\!+\!\omega_{\mk 2})$ is a factor of the Fourier
transform of source emission time distribution, $\rho(\mr)$ is source
density distribution.

As an expansion of the spherical sources \cite{{Padula06,Padula10,
Padula10a,YZHANG14}}, we take here the source density distribution
and velocity as
\begin{equation}
\label{rdis}
\rho(\mr)=C e^{-(\mx^2+\my^2)/(2R_T^2)}e^{-\mz^2/(2R_L^2)}\,
\theta(\sqrt{\mx^2+\my^2}-2R_T)\,\theta(|\mz|-2R_L),
\end{equation}
\begin{equation}
\mv(\mr)=\left[a_x\mx/(2R_T),a_y\my/(2R_T),a_z\mz/(2R_L)\right],
\end{equation}
where $C$ is a normalization constant, $R_T$ and $R_L$ are the Gaussian
radii of the source in the transverse and longitudinal directions, $a_x$,
$a_y$, and $a_z$ are three velocity parameters.
The emission time distribution is taken to be the typical exponential decay \cite{AsaCsoGyu99,Padula06,Padula10,Padula10a,YZHANG14},
\begin{equation}
F(\tau)=\frac{\theta(\tau-\tau_0)}{\Delta t}\;e^{-(\tau-\tau_0)/\Delta t},
\end{equation}
and its Fourier transform square is $|{\widetilde F}(\omega_\mk+\omega_{-\mk})
|^2=[1+(\omega_\mk+\omega_{-\mk})^2\Delta t^2]^{-1}$.
In the calculations, we take the parameter $\Delta t$ to be 2 fm/$c$ as
in \cite{Padula06,Padula10,Padula10a,YZHANG14}, and take $R_T=5$ fm, $R_L
=4$ fm.  The chemical potential of boson-antiboson pairs, $\mu_{1,2}
(\mr)$, is taken to be zero, and $T(\mr)$ is taken as fixed freeze-out
temperatures, approximately, $T_f=140$ MeV for $\phi$ meson and $T_f=170$
MeV for $K$ meson \cite{Padula06,Padula10,Padula10a,YZHANG14}, respectively.

\begin{figure}[htbp]
\includegraphics[scale=0.6]{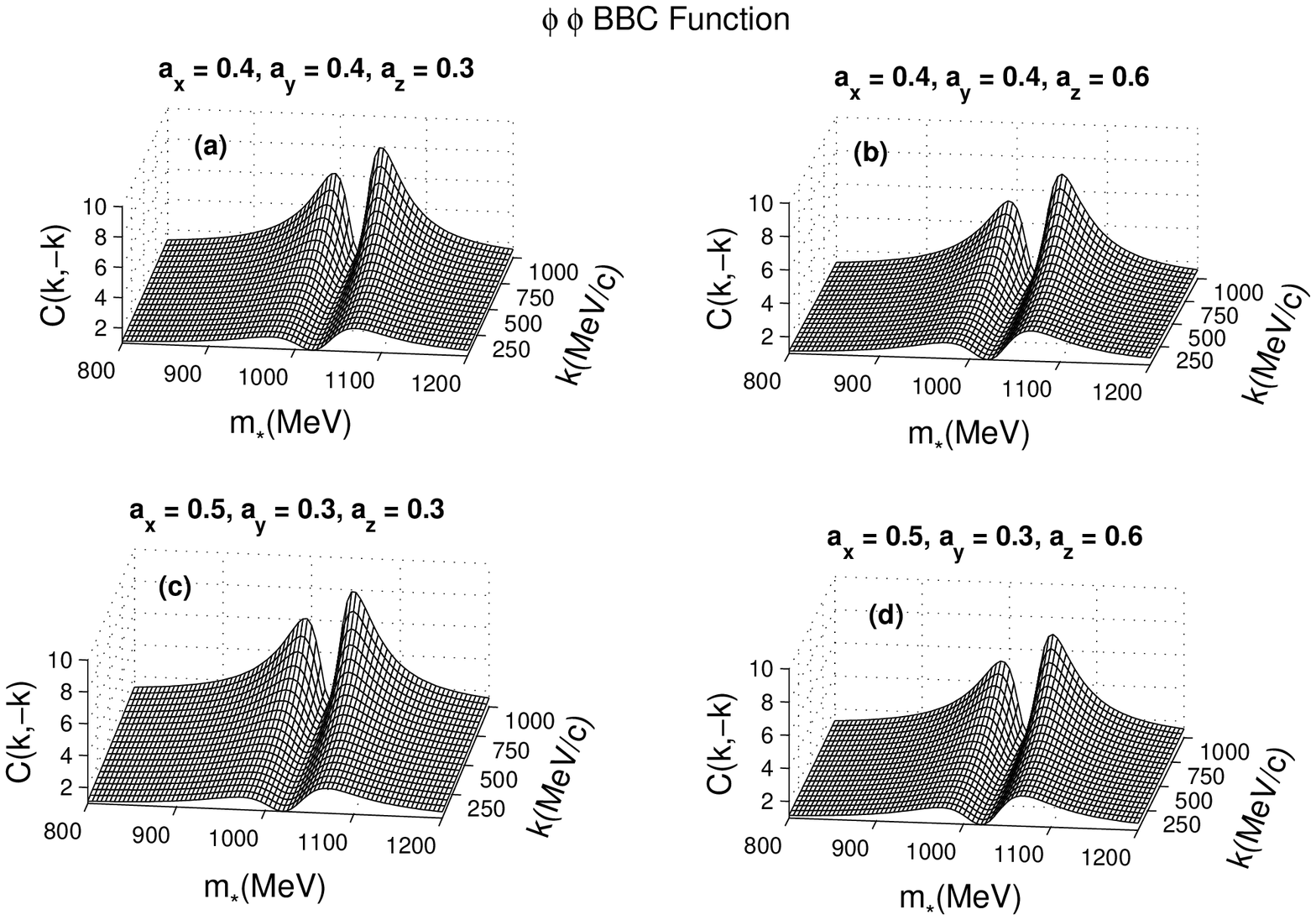}
\vspace*{-5mm}
\caption{The BBC functions of $\phi\,\phi$ in $m_*$-$k$ plane for the
anisotropic sources. }
\label{BBCFphi}
\end{figure}

We show in Fig. \ref{BBCFphi} the BBC functions of $\phi\,\phi$ in
$m_{\!*}$-$k$ plane for the anisotropic sources with different velocity
parameters.  For a fixed mass shift, the BBC functions increase with
momentum magnitude of particle.  The peaks of the BBC functions for the
sources with smaller longitudinal velocity are higher than those with
larger longitudinal velocity.  The pattern of BBC function for $a_x=a_y
=0.4$ is almost the same as that for $a_x=0.5$ and $a_y=0.3$, because
the two sources have almost the same transverse velocity.

\begin{figure}[htbp]
\includegraphics[scale=0.6]{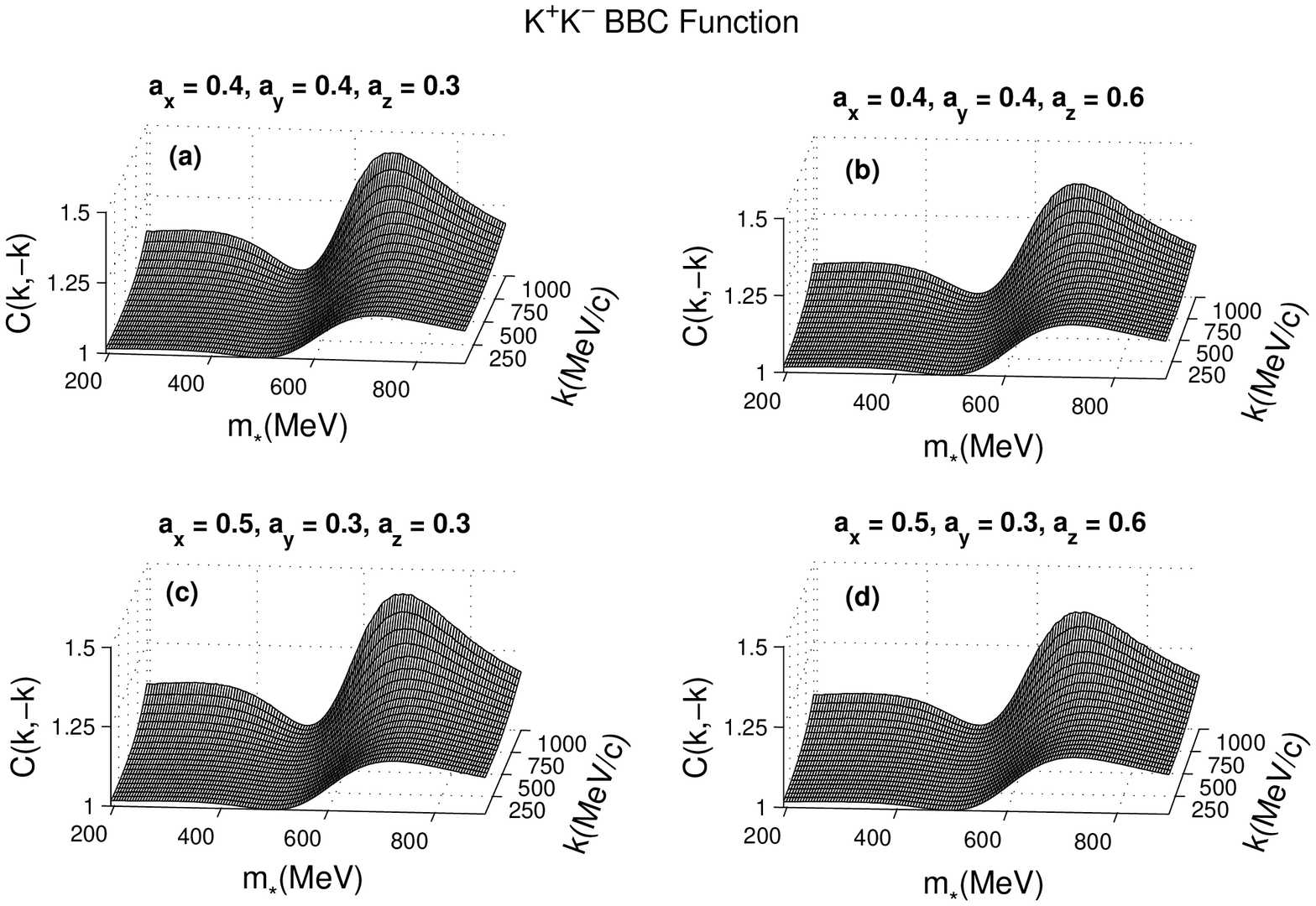}
\vspace*{-5mm}
\caption{The BBC functions of $K^+K^-$ in $m_*$-$k$ plane for the
anisotropic sources. }
\label{BBCFK}
\end{figure}

In Fig. \ref{BBCFK},  we plot the BBC functions of $K^+K^-$ for the
sources with different velocity parameters.  As compared with the BBC
functions of $\phi\, \phi$, the BBC functions of $K^+K^-$ are wider
in $m_{\!*}$ distribution, and have lower peak values.

\section{The dependence of BBC function on particle momentum direction}
\begin{figure}[htbp]
\includegraphics[scale=0.6]{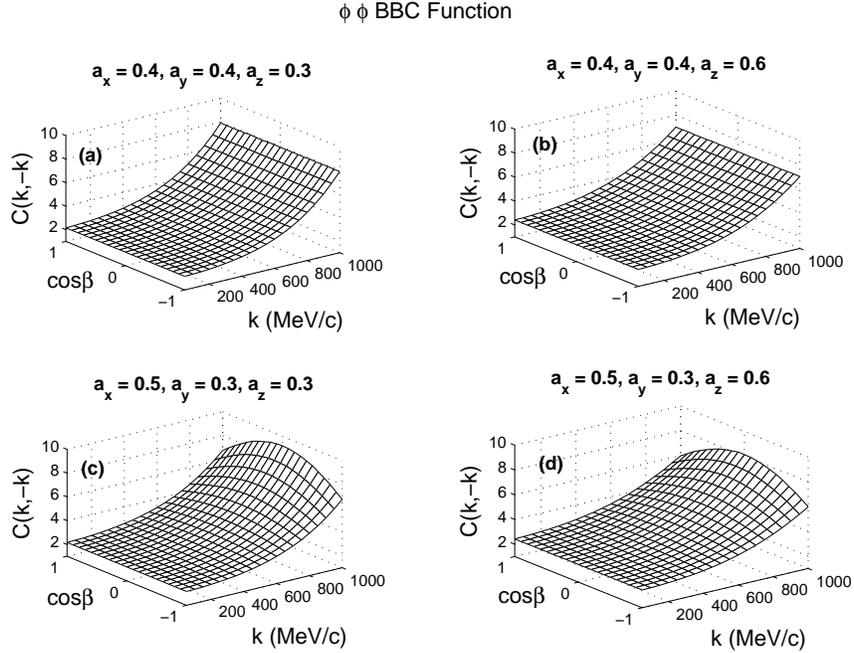}
\vspace*{-5mm}
\caption{The BBC functions of $\phi\,\phi$ in $\cos\beta$-$k$ plane for
the anisotropic sources. $m_*=1050$ MeV. }
\label{BBCcosbp}
\end{figure}

For anisotropic sources, the BBC functions depend not only on the
magnitude of particle momentum, but also on its direction.  We introduce
\begin{equation}
\cos\alpha=k_z/|{\mk}|,~~~~~\cos\beta=k_x/|\mk_T|,~~~~
\left(|\mk_T|=\sqrt{k_x^2+k_y^2}\,\right),
\end{equation}
to describe the direction of particle momentum.  The variation of the
BBC functions of $\phi\,\phi$ with $\cos\beta$ for the anisotropic
sources are shown in Fig. \ref{BBCcosbp}, where $m_{\!*}$ is taken as
1050 MeV corresponding approximately to the peaks of the BBC functions
in Fig. \ref{BBCFphi}.  For the sources with isotropic transverse
velocity ($a_x =a_y$), the BBC functions are independent of $\cos\beta$,
and the values of the BBC functions at large $k$ decrease slightly with
increasing longitudinal velocity.
The BBC functions for the sources with anisotropic transverse velocity
($a_x>a_y$) show an obvious dependence of transverse momentum direction.
For the sources with $a_x>a_y$, the average velocities $\langle v_x\rangle
> \langle v_y\rangle$, and the values of the BBC functions at large $k$
reach maximums/minimums when the transverse momentum is parallel with the
smaller/larger transverse velocity ($v_y$/$v_x$) direction.

\begin{figure}[htbp]
\includegraphics[scale=0.6]{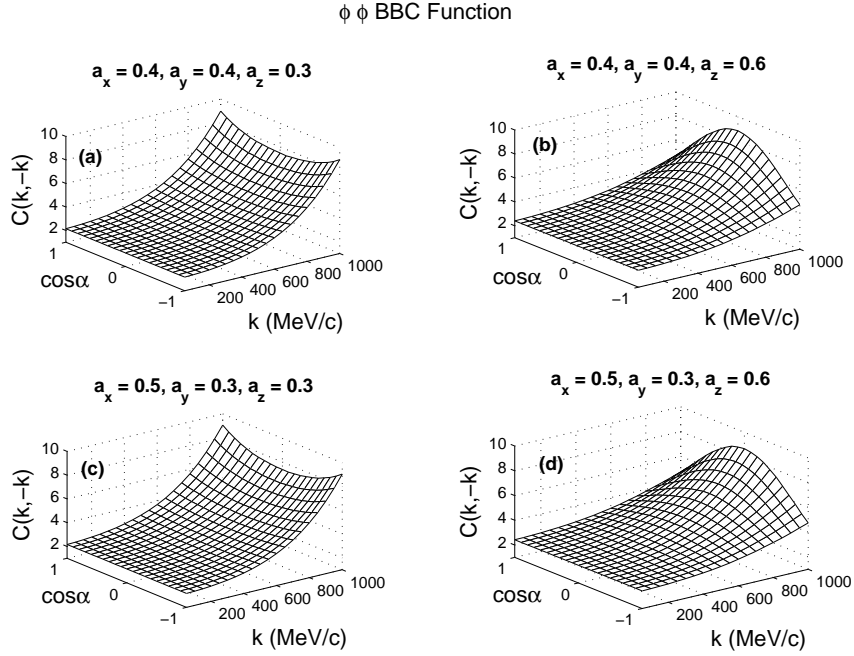}
\vspace*{-5mm}
\caption{The BBC functions of $\phi\,\phi$ in $\cos\alpha$-$k$ plane for
the anisotropic sources. $m_*=1050$ MeV. }
\label{BBCcosap}
\end{figure}

We plot in Fig. \ref{BBCcosap} the BBC functions of $\phi\,\phi$ in $\cos
\alpha$-$k$ plane for the anisotropic sources with the same parameters as
Fig. \ref{BBCcosbp}.  For the sources with $a_z=0.3$, the average transverse
velocity is larger than the average longitudinal velocity, $\langle v_T
\rangle > \langle v_z\rangle$.  The values of the BBC functions at large
$k$ for the sources with $a_z=0.3$ have maximums/minimums when particle
momentum is parallel with the smaller/larger velocity ($v_z$/$v_T$) direction.
On the other hand, there is $\langle v_z\rangle > \langle v_T\rangle$ for
the sources with $a_z=0.6$.  In this case the values of the BBC functions
at large $k$ have maximums/minimums when particle momentum is parallel with
the smaller/larger velocity ($v_T$/$v_z$) direction.

\begin{figure}[htbp]
\includegraphics[scale=0.6]{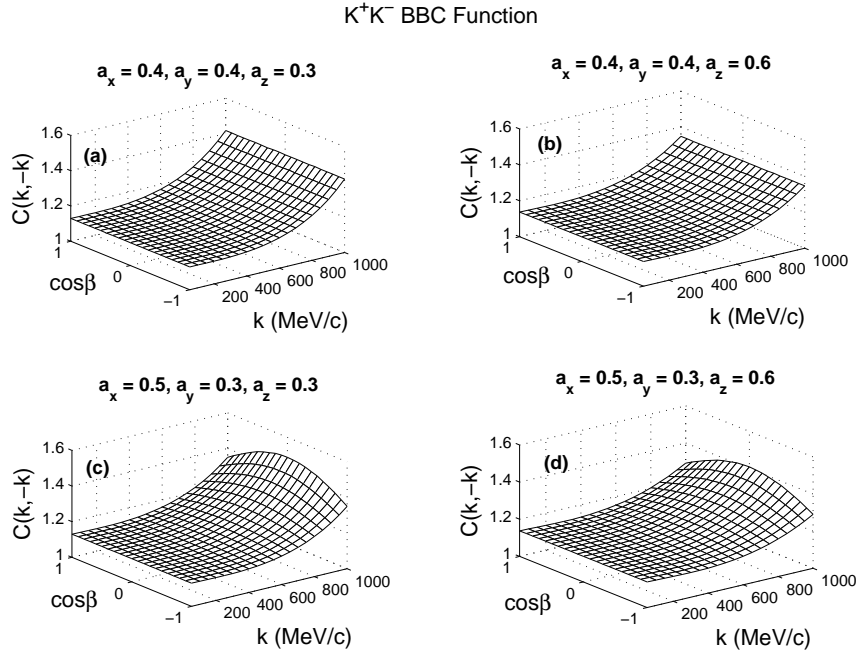}
\vspace*{-5mm}
\caption{The BBC functions of $K^+K^-$ in $\cos\beta$-$k$ plane for the
anisotropic sources. $m_*=650$ MeV. }
\label{BBCcosbk}
\end{figure}

\begin{figure}[htbp]
\includegraphics[scale=0.6]{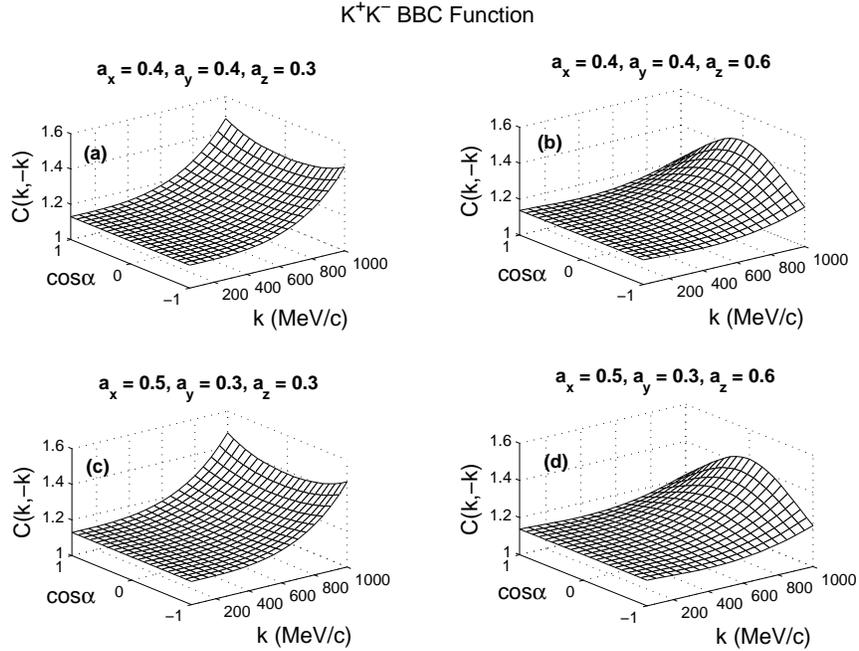}
\vspace*{-5mm}
\caption{The BBC functions of $K^+K^-$ in $\cos\alpha$-$k$ plane for the
anisotropic sources. $m_*=650$ MeV. }
\label{BBCcosak}
\end{figure}

In Figs. \ref{BBCcosbk} and \ref{BBCcosak}, we plot the BBC functions of
$K^+K^-$ for the anisotropic sources, in $\cos\beta$-$k$ and $\cos\alpha$-$k$
planes, respectively.  Here $m_{\!*}$ is taken as 650 MeV corresponding
approximately to the peaks of the BBC functions in Fig. \ref{BBCFK}.
One can see that the BBC functions have the similar dependence of particle
momentum direction as that of $\phi\,\phi$ BBC functions.

\section{The effect of source velocity on BBC functions}
We have seen that the BBC functions are sensitive to the magnitude and direction
of particle momentum for the anisotropic source.  Next, we further examine the
effect of source velocity on the BBC functions.  From Eqs. (\ref{Gc}) ---
(\ref{nkk}), one sees that the BBC functions will increase when $n_{\mk,-\mk}$
decreases \cite{AsaCsoGyu99}.  Because $n_{\mk,-\mk}$ is related to $k^{\mu}
u_{\mu}$, the effect of source velocity on the BBC function will vary not only
with the magnitude of the velocity, but also with the angle, $\theta$, between
the velocity and particle momentum.

\begin{figure}[htbp]
\includegraphics[scale=0.6]{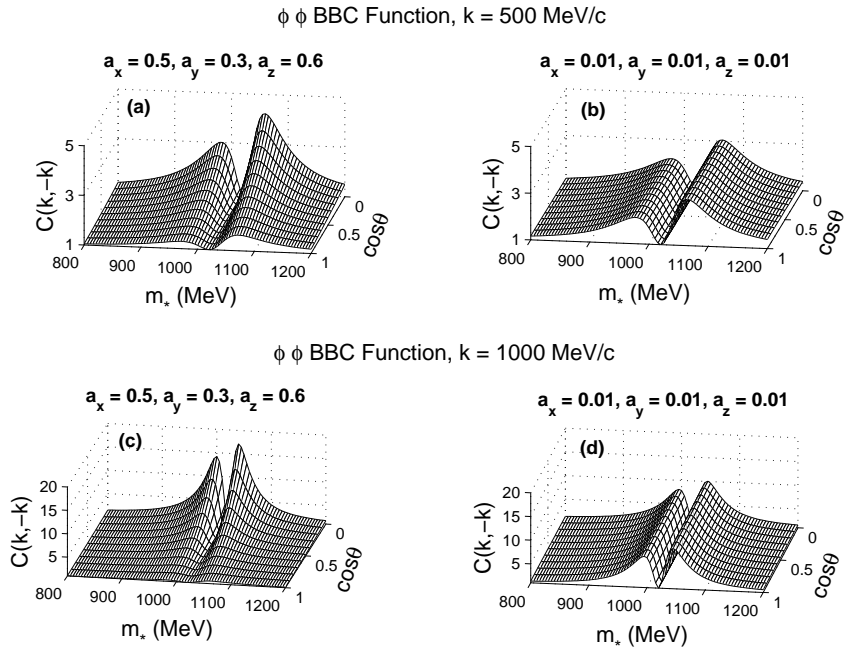}
\vspace*{-5mm}
\caption{The BBC functions of $\phi\,\phi$ in $m_*$-$\cos\theta$ plane
for the sources with the anisotropic velocity and almost zero velocity.
(a) and (b) $k=500$ MeV/$c$; (c) and (d) $k=1000$ MeV/$c$. }
\label{BBCcostm}
\end{figure}

\begin{figure}[htbp]
\vspace*{-5mm}
\includegraphics[scale=0.6]{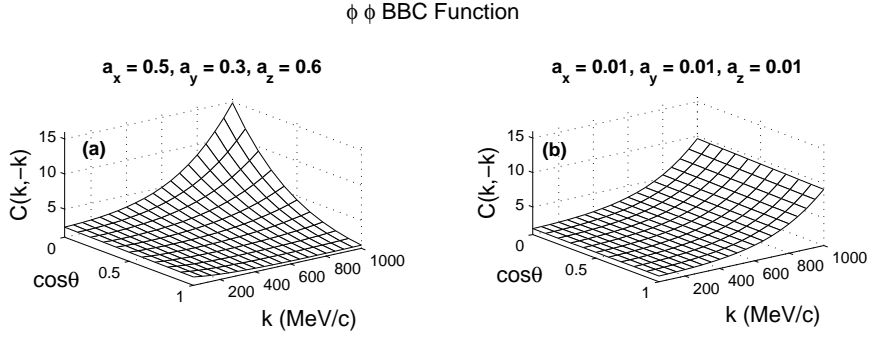}
\vspace*{-30mm}
\caption{The BBC functions of $\phi\,\phi$ in $\cos\alpha$-$k$ plane
for the sources with the anisotropic velocity and almost zero velocity.
$m_*=1050$ MeV. }
\label{BBCcostk}
\end{figure}

We plot in Fig. \ref{BBCcostm} the BBC functions of $\phi\,\phi$ in
$m_{\!*}$-$\cos\theta$ plane for the source with the anisotropic velocity
[panels (a) and (c)] and the source with almost zero velocity [panels (b)
and (d)], for the fixed $k=500$ and 1000 MeV/$c$.  The effect of velocity
leads to the smaller/larger values of the BBC functions at large/smaller
$\cos\theta$, as compared to the approximately static source.  In Fig.
\ref{BBCcostk}, we plot the BBC functions of $\phi\,\phi$ in $\cos\theta$-
$k$ plane for the sources with the anisotropic velocity and almost zero
velocity.  For the anisotropic expanding source, The BBC function at a large
$k$ and small $\cos\theta$ is larger than that for the almost static source,
and the BBC functions at larger $k$ decrease with $\cos\theta$ rapidly.

\begin{figure}[htbp]
\includegraphics[scale=0.6]{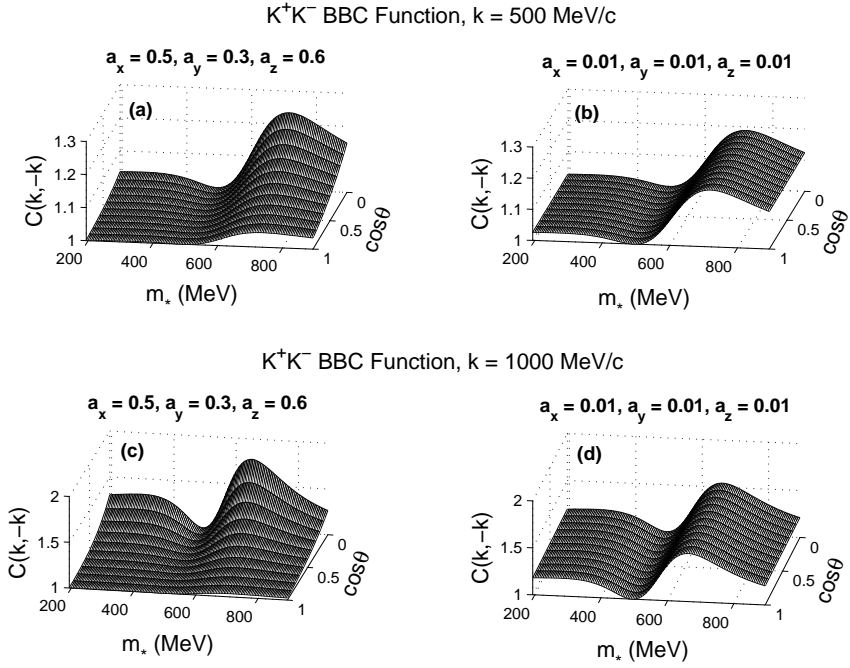}
\vspace*{-5mm}
\caption{The BBC functions of $K^+K^-$ in $m_*$-$\cos\theta$ plane for
the sources with the anisotropic velocity and almost zero velocity.
(a) and (b) $k=500$ MeV/$c$; (c) and (d) $k=1000$ MeV/$c$. }
\label{BBCcostmk}
\end{figure}

\begin{figure}[htbp]
\vspace*{-5mm}
\includegraphics[scale=0.6]{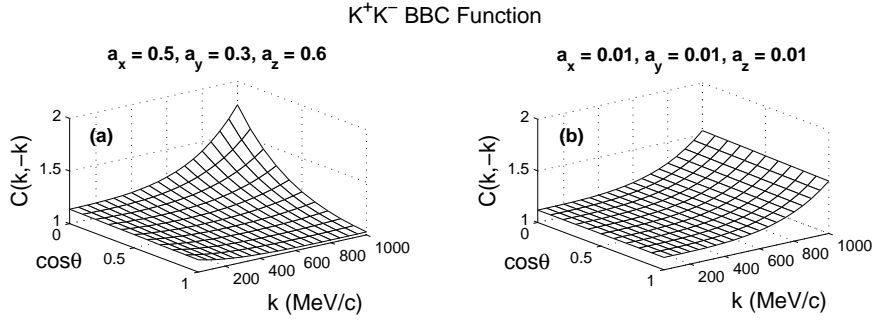}
\vspace*{-30mm}
\caption{The BBC functions of $K^+K^-$ in $\cos\theta$-$k$ plane for the
sources with the anisotropic velocity and almost zero velocity.  $m_*=650$
MeV. }
\label{BBCcostkk}
\end{figure}

In Figs. \ref{BBCcostmk} and \ref{BBCcostkk}, we plot the BBC functions
of $K^+K^-$ in $m_{\!*}$-$\cos\theta$ and $\cos\theta$-$k$ planes for the
sources with the anisotropic velocity and almost zero velocity.
The effect of the source velocity on the BBC functions are similar to
that on the BBC functions of $\phi\,\phi$, but a little smaller.

\begin{figure}[htbp]
\vspace*{3mm}
\includegraphics[scale=0.75]{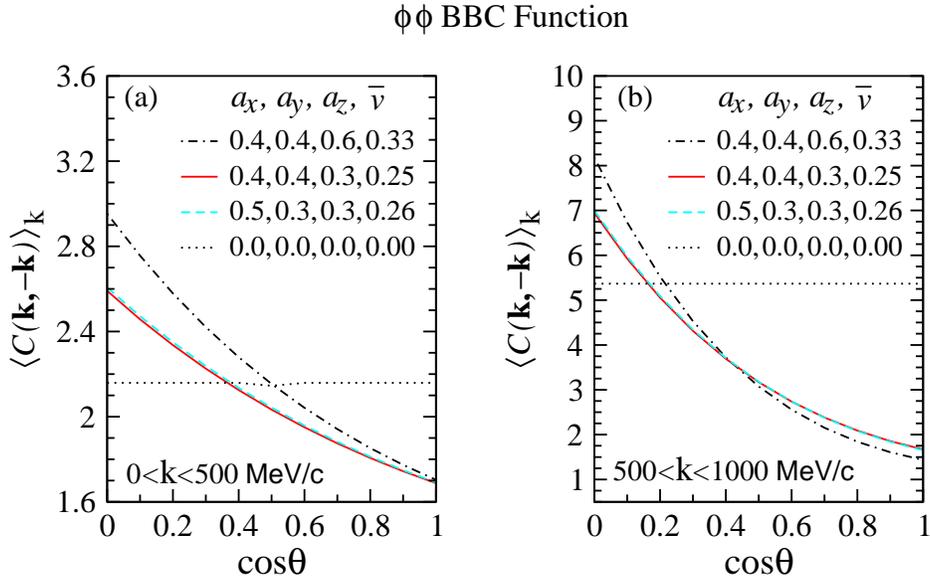}
\caption{The $\cos\theta$-dependence of the average BBC functions of
$\phi\,\phi$ over particle momentum regions (a) $0<k<500$ MeV/$c$ and
(b) $500<k<1000$ MeV/$c$, for the sources with the different values of
the velocity parameters, $a_x$, $a_y$, $a_z$, and the corresponding
average source velocity, $\bar v$.  Here, $m_*=1050$ MeV. }
\label{pBBCcost2d}
\end{figure}

In Fig. \ref{pBBCcost2d}, we show the dependences of the averaged BBC
functions of $\phi\,\phi$ on the angle between the particle momentum
and source velocity, for the sources with the different values of the
velocity parameters, $a_x$, $a_y$, $a_z$, and the corresponding average
source velocity, $\bar v$.  Here, $m_*$ is taken as 1050 MeV, the
momentum regions are $0<k<500$ MeV/$c$ and $500<k<1000$ MeV/$c$ for
Figs. \ref{pBBCcost2d}(a) and \ref{pBBCcost2d}(b), respectively.  By
comparing the results of the BBC functions for the expanding sources
and the static source, one sees that as the source velocity increases,
the average BBC function increases when the particle momentum is
perpendicular to the velocity, and the average BBC function decreases
when the particle momentum is nearly parallel to the velocity.  The
BBC function is greatest when the particle momentum is perpendicular
to the source velocity at $\cos \theta=0$, and the BBC is smallest
when the particle momentum is parallel to the source velocity at
$\cos \theta$=1.

\begin{figure}[htbp]
\vspace*{5mm}
\includegraphics[scale=0.75]{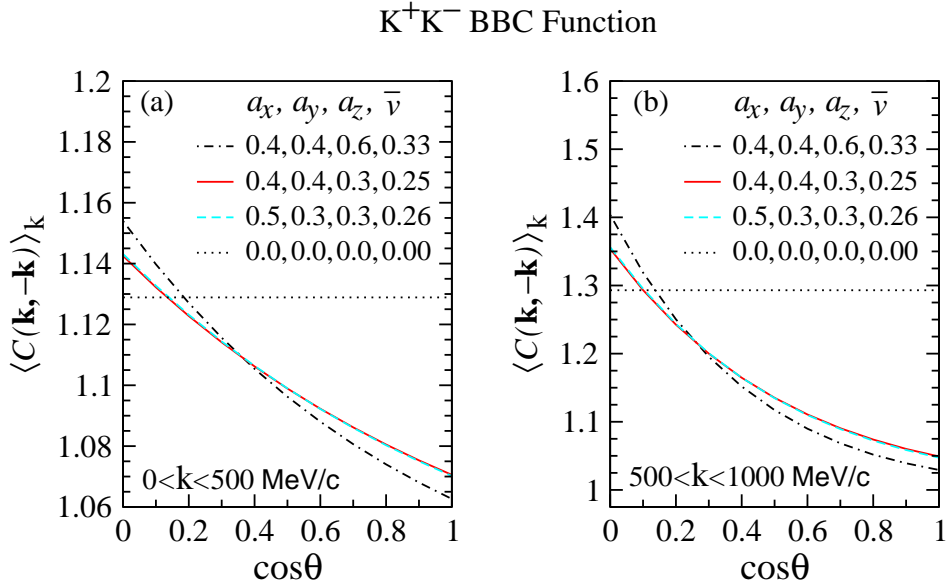}
\vspace*{0mm}
\caption{The $\cos\theta$-dependence of the average BBC functions of
$K^+K^-$ over particle momentum regions (a) $0<k<500$ MeV/$c$ and (b)
$500<k<1000$ MeV/$c$, for the sources with the different values of
the velocity parameters, $a_x$, $a_y$, $a_z$, and the corresponding
average source velocity, $\bar v$.  Here, $m_*=650$ MeV. }
\label{kBBCcost2d}
\end{figure}

In Fig. \ref{kBBCcost2d}, we show the dependences of the averaged BBC
functions of $K^+K^-$ on the angle between the particle momentum and
source velocity, for the sources with the different values of the
velocity parameters, $a_x$, $a_y$, $a_z$, and the corresponding
average source velocity, $\bar v$.  Here, $m_*$ is taken as 1050 MeV,
the momentum regions are $0<k<500$ MeV/$c$ and $500<k<1000$ MeV/$c$
for Figs. \ref{kBBCcost2d}(a) and \ref{kBBCcost2d}(b), respectively.
At $\cos \theta \sim 0$, the average BBC function increases with
increasing source expansion velocity.  While at $\cos\theta\sim 1$,
the average BBC function decreases with the increasing source expansion
velocity.

\section{Summary and conclusion}
As an extension of the previous works of the BBC functions for spherical
hadronic expanding sources \cite{Padula06,Padula10,Padula10a,YZHANG14},
we calculate the BBC functions of $\phi\,\phi$ and $K^+K^-$ for the
anisotropic expanding sources.  As compared to the spherical source
models, the anisotropic sources are a more realistic case for the hadronic
sources formed in high energy heavy ion collisions, and the investigations
of the BBC functions for the anisotropic sources may provide additional
signals for experimental detection.  The values of the BBC functions for
the expanding sources depend not only on the magnitude of particle momentum,
but also on its direction.  For the sources with the average transverse
velocities, $\langle v_x \rangle >\langle v_y\rangle$, the values of the
BBC functions at large particle momentum reach maximums (or minimums) when
the transverse momentum is parallel with the smaller (or larger) velocity
direction.  For the sources with $\langle v_z \rangle <\langle v_T \rangle$
[or $\langle v_z \rangle >\langle v_T \rangle$], the values of the BBC
functions at large particle momentum have maximums (or minimums) when the
momentum is parallel with $\mv_z$ (or $\mv_T$).  We further investigate
the effect of the anisotropic source velocity on the BBC functions.
As the source expansion velocity increases, the BBC function increases
when the particle momentum is perpendicular to the source velocity, and
the BBC function decreases when the particle momentum is parallel to the
source velocity.

\begin{acknowledgments}
This research was supported by the National Natural Science Foundation of China under Grant No.
11275037.
\end{acknowledgments}

\end{document}